\documentclass[aps,prd,letterpaper,superscriptaddress,groupedaddress,preprintnumbers,floatfix,twocolumn,nofootinbib]{revtex4}  
\usepackage{graphicx,comment}  
\usepackage{dcolumn}   
\usepackage{bm}        
\usepackage{amssymb}   

\usepackage{subfigure}
\usepackage{multirow}
\usepackage{placeins}
\usepackage{epstopdf}
\usepackage{mathtools}
\usepackage{color}

\usepackage{amsfonts}    
\usepackage{amsmath}
\usepackage{dcolumn}

\newcolumntype{d}[1]{D{.}{.}{1.8}}

\begin{document}

\preprint{
\vbox{
\hbox{MIT-CTP/5071}
}}

\title{The pressure distribution and shear forces inside the proton}

\author{P.~E.~Shanahan}\affiliation{Center for Theoretical Physics, Massachusetts Institute of Technology, Cambridge, MA 02139, U.S.A.}
\affiliation{Perimeter Institute for Theoretical Physics, Waterloo, Ontario N2L 2Y5, Canada}
\author{W.~Detmold}      \affiliation{Center for Theoretical Physics, Massachusetts Institute of Technology, Cambridge, MA 02139, U.S.A.}
	
\begin{abstract}
The distributions of pressure and shear forces inside the proton are investigated using lattice Quantum Chromodynamics (LQCD) calculations of the energy momentum tensor, allowing the first model-independent determination of these fundamental aspects of proton structure. This is achieved by combining recent LQCD results for the gluon contributions to the energy momentum tensor 
with earlier calculations of the quark contributions. The utility of LQCD calculations in exploring, and supplementing, the assumptions in a recent extraction of the pressure distribution in the proton from deeply virtual Compton scattering is also discussed. 
Based on this study, the target kinematics for experiments aiming to determine the pressure and shear distributions with greater precision at Thomas Jefferson National Accelerator Facility and a future Electron Ion Collider are investigated.
\end{abstract}

\maketitle

Many of the most fundamental aspects of hadron structure are encoded in form factors that describe the hadron's interactions with the electromagnetic, weak, and gravitational forces. In the forward limit, the electromagnetic form factors reduce to the electric charge and magnetic moment of a hadron, weak form factors reduce to the axial charge and induced pseudoscalar coupling, while the gravitational form factors describe the hadron's mass, spin, and $D$-term. Unlike the mass, spin, and electromagnetic and weak properties of the proton, which are well-known, the quark $D$-term form factor, $D_{q}(t,\mu)$ (where $t$ is the squared momentum transfer and $\mu$ is a renormalisation scale), has only recently been extracted from experiment~\cite{Burkert:2018bqq}. The gluon term $D_{g}(t, \mu)$ has never been extracted. These functions, which parameterise the spatial-spatial components of the energy momentum tensor (EMT),  describe the internal dynamics of the system through the pressure and shear distributions of the proton \cite{Polyakov:2002wz,Polyakov:2002yz,Polyakov:2018zvc}.

While the quark and gluon contributions to the pressure distribution are scale- and scheme-dependent and depend on the non-conserved components of the EMT \cite{Polyakov:2018zvc}, the  
total pressure distribution (the sum of the quark and gluon contributions) is a measurable quantity defined purely from the $D$-term. As such, it is of fundamental interest as one of the few remaining aspects of proton structure about which very little is known.
Recently, the pressure distribution in the proton was extracted for the first time from deeply virtual Compton scattering (DVCS) experiments at the Thomas Jefferson National Accelerator Facility (JLab) by Burkert, Elhouadrhiri and Girod  \cite{Burkert:2018bqq} (henceforth referred to as BEG) over a limited kinematic range. The result is remarkable; it indicates that the internal pressure in a proton is approximately $10^{35}$ Pascal, exceeding the estimated pressure in the interior of a neutron star. However, since DVCS is almost insensitive to gluons, this determination necessarily relies on several assumptions about the gluon contributions to the proton pressure that are important to investigate.

This letter presents the first determination of the QCD pressure and shear distributions inside the proton, including both the quark and gluon contributions to these quantities. The study is undertaken using lattice Quantum Chromodynamics (LQCD) with larger-than-physical values of the light quark masses. 
The results reveal that gluons play an important role, different from that of quarks, in the proton's internal dynamics. In particular, the gluon contribution to the $D$-term form factor, which dictates the pressure and shear distributions, is distinguished in both magnitude and momentum-dependence from the quark contribution. At the scale $\mu=2$~GeV in the $\overline{\text{MS}}$ scheme, gluons provide the dominant contributions to the proton shear distribution (for which the separation is well-defined). 
The utility of these LQCD results in augmenting the experimental extraction of the pressure in BEG is also explored. While the calculations provide some support to the assumptions made in that pioneering work, they also indicate deficiencies that must be remedied before a completely model-independent determination of the pressure and shear distributions is possible from experiment. Based on the LQCD studies, the kinematics of future experiments at JLab, a future Electron Ion Collider (EIC), and other facilities that will be needed to achieve this are discussed.\\[-6pt]

{\bf The EMT and $D$-term form factors}:  The pressure and shear distributions in the proton are constructed from the $D$-term form factors, which are defined from the nucleon matrix elements of the traceless, symmetric energy-momentum tensor. Precisely, the matrix elements of the gluon component of the EMT,
\begin{eqnarray}\label{Eq:EMT-FFs-spin-12}
\langle p^\prime,s^\prime| G^a_{\alpha\{\mu}
G^{a\alpha}_{\nu\}} |p,s\rangle
= \bar u'{\cal F}_{\mu\nu}[A_g,B_g,D_g] u
 \hspace*{1.75cm}
\\ 
=\bar u'\Big[ A_g\,\gamma_{\{\mu} P_{\nu\}}
+ B_g\,\frac{i\,P_{\{\mu}\sigma_{\nu\}\rho}\Delta^\rho}{2M_N}
+ D_g\,\frac{\Delta_{\{\mu}\Delta_{\nu\}}}{4M_N} \Big] u\,,
\nonumber
\end{eqnarray}
depend on three generalised form factors (GFFs), $A_g(t,\mu)$, $B_g(t,\mu)$ and $D_g(t,\mu)$, that are 
functions of the momentum transfer $t=\Delta^2$ with $\Delta_\mu=p'_{\mu}-p_\mu$.  In Eq.~\eqref{Eq:EMT-FFs-spin-12}, $G_{\mu\nu}^a$ is the gluon field strength tensor, braces denote symmetrisation and trace-subtraction of the enclosed indices, $P_\mu =(p_\mu+p_\mu')/2$, the spinors are expressed as $u=u_s(p)$ and $\overline{u}'=\overline{u}_{s'}(p')$, and $M_N$ is the proton mass. An exactly analogous decomposition exists for matrix elements of the quark  EMT:
\begin{align}
\langle p^\prime,s^\prime|  \overline\psi_q \gamma_{\{\mu} i\overset{\text{\tiny$\leftrightarrow$}}{D}_{\nu\}}\psi_q|p,s\rangle
= \bar u' {\cal F}_{\mu\nu}[A_q,B_q,D_q] u,
\label{Eq:EMT-FFs-spin-12-quark}
\end{align}
where $\psi_q$ is the quark field of flavour $q$ and ${D}_{\nu}$ is the gauge covariant derivative.

The individual EMT form factors depend on the renormalisation scheme and scale, $\mu$.
Since the isoscalar combinations of twist-two operators in Eqs.~\eqref{Eq:EMT-FFs-spin-12} and \eqref{Eq:EMT-FFs-spin-12-quark}  
mix under renormalisation, so too do the individual isoscalar quark and gluon form factors. This mixing takes the form
\begin{eqnarray}
\begin{pmatrix}
D_{u+d}(t,\mu)
\\ D_g(t,\mu)
\end{pmatrix}
= 
\begin{pmatrix}
Z_{qq}(\frac{\mu}{\mu'} )& Z_{qg}(\frac{\mu}{\mu'} ) \\
Z_{gq}(\frac{\mu}{\mu'}) & Z_{gg}(\frac{\mu}{\mu'} )
\end{pmatrix}
\begin{pmatrix}
D_{u+d}(t,\mu')
\\ D_g(t,\mu')
\end{pmatrix},
\label{eq:evolve}
\end{eqnarray}
where the perturbative mixing coefficients are given in Ref.~\cite{Yang:2016xsb}.
Because of conservation of the EMT, the isoscalar combination of the quark and gluon pieces, $D(t) = D_{u+d}(t,\mu)+ D_g(t,\mu)$, is scale-invariant.

In terms of the total $D(t)$ form factor, the  shear and pressure distributions in the proton can be expressed in the Breit frame as~\cite{Polyakov:2002wz,Polyakov:2002yz,Polyakov:2018zvc}
\begin{equation}
\label{Eq:relationSPD}
s(r)= -\frac{r}{2} \frac{d}{dr} \frac{1}{r} \frac{d}{dr}
{\widetilde{D}(r)}, \quad
p(r)=\frac{1}{3} \frac{1}{r^2}\frac{d}{dr} r^2\frac{d}{dr}
{\widetilde{D}(r)},
\end{equation}
respectively, where 
\begin{equation}
{\widetilde{D}(r)=}
\int {\frac{d^3\vec{p}}{2E(2\pi)^3}}\ e^{{-i} {\vec{p}\cdot\vec{r}}}\ D(-{\vec{p}\,}^2).
\end{equation}
While scale-dependent quark and gluon contributions to the shear forces $s_a(r,\mu)$ can be computed from the $D_{q,g}(t,\mu)$, only the total pressure distribution $p(r)$ can be determined; the individual quark and gluon contributions to the pressure distribution depend not only on the $D$-term GFFs but also on additional GFFs related to the  trace  of the EMT (that cancel in the sum \cite{Polyakov:2018zvc}).\\[-8pt]

\begin{figure}[!t]
	\includegraphics[width=\columnwidth]{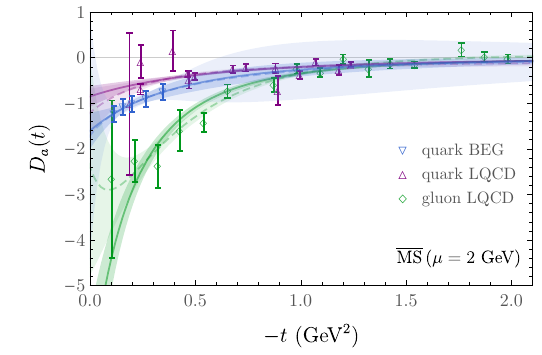}
	\caption{LQCD calculations of $D_{u+d}^{\text{(conn.)}}(t,\mu)$ (purple triangles)~\cite{Hagler:2007xi} and $D_g(t,\mu)$ (green diamonds)~\cite{Shanahan:2018pib} at the scale $\mu=2$ GeV in the $\overline{\text{MS}}$ scheme. The BEG extracted $D$-term (blue inverted triangles), rescaled to $\mu=2$ GeV, is also shown for comparison. The shaded bands denote tripole (solid) and modified $z$-expansion (dashed, Eq.~\eqref{eq:modz}) fits to each data set.
	}
	\label{fig:comparison}
\end{figure}

{\bf Lattice QCD quark and gluon $D$-term form factors}:  The quark GFFs of the proton have been computed 
by a number of LQCD collaborations \cite{Syritsyn:2011vk,Bali:2013dpa,Hagler:2007xi,Bratt:2010jn,Sternbeck:2012rw,Brommel:2007sb} since the first studies in Refs.~\cite{Hagler:2003jd,Gockeler:2003jfa} (see  Ref.~\cite{Hagler:2009ni} for a review). While there are as-yet no calculations directly at the physical quark masses, studies over masses corresponding to $0.21\le m_\pi \alt 1.0$ GeV show very mild mass-dependence relative to the other statistical and systematic uncertainties of the calculations. The $t$-dependence of the GFFs  has been determined over the range $0\le|t|\alt 2$ GeV$^2$. The calculations contain all contributions for the isovector combination $D_{u-d}(t,\mu)$, while so-called disconnected contractions have been neglected in existing determinations of the isoscalar quark GFFs, $D_{u+d}(t,\mu)$, since these terms are both particularly numerically challenging to compute and are found to be small for many other quantities. An important observation from these determinations of the GFFs is that the isovector combination $D_{u-d}(t,\mu)\sim0$ over the entire range of quark masses and momentum transfers that have been studied. An example of the isoscalar connected quark $D$-term form factor from Ref.~\cite{Hagler:2007xi} is shown in Fig.~\ref{fig:comparison} at quark masses corresponding to $m_\pi\sim 0.5$ GeV.

The gluon $D$-term form factor  was recently determined for the first time in Ref.~\cite{Shanahan:2018pib} at a single set of  quark masses corresponding to $m_\pi\sim0.45$ GeV and at a single lattice spacing and volume.\footnote{Note that these LQCD calculations of the gluon $D$-term differ in their action, lattice spacing, quark masses, and lattice volume from the quark calculations of Ref.~\cite{Hagler:2007xi}. Both the quark and gluon calculations are subject to systematic uncertainties arising from the fixed parameters of the simulations; it is expected that uncertainties from the finite lattice spacing are of ${\cal O}(a \Lambda_{\rm QCD})\sim10$\%, except at  $|t|\ge1$ GeV$^2$ where contribution of ${\cal O}(a \sqrt{t})\alt50$\% may arise (although note that the form factors are small relative to the statistical uncertainties in this regime). Finite volume effects are expected to be ${\cal O}(e^{-m_\pi L})\alt1$\%. 
} 
Mixing with the isoscalar quark operators in Eq.~\eqref{Eq:EMT-FFs-spin-12-quark} was  neglected based on perturbative arguments \cite{Alexandrou:2016ekb}. The uncertainties, which encompass statistical and systematic effects in the gluon $D$-term calculations, are somewhat larger than for the quark form factor because of  a more complicated renormalisation procedure and the much larger statistical variance of gluonic quantities. 
Based on chiral perturbation theory \cite{Belitsky:2002jp,Detmold:2005pt,Ando:2006sk,Diehl:2006ya,Dorati:2007bk}, the quark-mass dependence of this isoscalar, purely gluonic quantity is expected to be mild compared with that of many other observables such as the nucleon electromagnetic form factors. 
Compared with the LQCD determination of the isoscalar quark $D$-term form factor at similar quark masses, the gluon form factor is approximately twice as large, with a  different $t$-dependence, as shown in Fig.~\ref{fig:comparison} and discussed in Ref.~\cite{Shanahan:2018pib}. \\[-8pt]

{\bf Model dependence:}
Since the pressure and shear distributions in Eq.~\eqref{Eq:relationSPD} involve Fourier transforms of the $D$-term form factor, a functional form must be used to interpolate and extrapolate the data determined at discrete values of $t$ over a finite interval. In order that the Fourier transform converges, the form factor must fall off at large $|t|$ faster than $1/|t|$. As discussed in BEG, a tripole form, which has the asymptotic behaviour expected from helicity selection rules \cite{Lepage:1980fj}, is a natural ansatz. Fits using this form describe the LQCD results reasonably well over their entire kinematic range, as shown in Fig.~\ref{fig:comparison}. Nevertheless, pressure and shear distributions determined under the assumption of this form suffer significant model-dependence, since there is no a-priori reason that $D(t)$ has such a simple form; it  need not be monotonic, nor positive definite.

An alternative parametrisation of the $t$-dependence of GFFs is provided by a modified $z$-expansion:
\begin{equation}\label{eq:modz}
D_{q/g}(t,\mu)=\frac{1}{(1-t/\Lambda^2)^3}\sum_{k=0}^{k_\text{max}}a_k \left[z(t)\right]^k,
\end{equation}
with 
$z(t) =[\sqrt{t_\text{cut}-t}-\sqrt{t_\text{cut}-t_0}]/[\sqrt{t_\text{cut}-t}+\sqrt{t_\text{cut}-t_0}]$. Since the conformal mapping guarantees analyticity around $z=0$, and unitarity guarantees convergence \cite{Hill:2010yb,Bourrely:2008za,Caprini:1997mu},  the $z$-expansion provides a more reliable estimate of uncertainties in regions unconstrained by data.
Modified $z$-expansion fits to the quark and gluon GFFs from LQCD, with the tripole mass $\Lambda$ fixed to that determined by a pure tripole fit to the GFF and with $k_\text{max}=2$, $t_\text{cut}=4m_\pi^2$, and $t_0=t_\text{cut}(1-\sqrt{1+(2\ \text{GeV})^2/t_\text{cut}})$, are shown in Fig.~\ref{fig:comparison}. In each case, the parametrisation is reasonably well constrained over a kinematic range that is sufficient for the GFFs to become indistinguishable from zero within uncertainties. Nevertheless, these fits are considerably less well constrained than the  tripole fits. 
Further discussion of the model-dependence in fits to the GFFs is provided in the Supplementary Material.\\[-6pt]

\begin{figure}
	\includegraphics[width=\columnwidth]{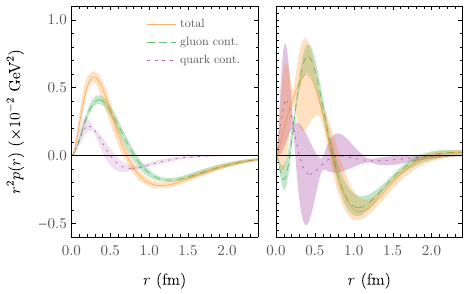}
	\caption{Left) Pressure distribution of the proton computed using tripole parameterisations of the LQCD quark $D$-term GFF and the LQCD gluon $D$-term GFF. The contributions from the quark and gluon terms are represented by the purple dotted and green dashed bands, respectively, while the total is denoted by the orange solid band.
		Right) The same quantities, determined based on modified $z$-expansion parametrisations of the $D$-term form factors.}
	\label{fig:tripolepressure}
\end{figure}

{\bf The pressure distribution and shear forces in the proton:} 
Figure~\ref{fig:tripolepressure} shows the pressure computed using the LQCD determinations of both quark and gluon $D$-term form factors for both the tripole parameterisation and modified $z$-expansion\footnote{To determine the error bands shown, a large number of re-samplings of the $D$-term form factors are fit with the appropriate functional form, and the pressure distribution is computed from each fit; the central value and error band of the displayed pressure distribution are the mean and standard deviation across those samplings at each value of $r$.}. Given the larger uncertainties in the latter fits to the $D$-term form factors, the $z$-expansion pressure is less well determined, although still resolved from zero by several standard deviations at the peak values. The differences provide an estimate of model-dependence.

\begin{figure}
	\includegraphics[width=\columnwidth]{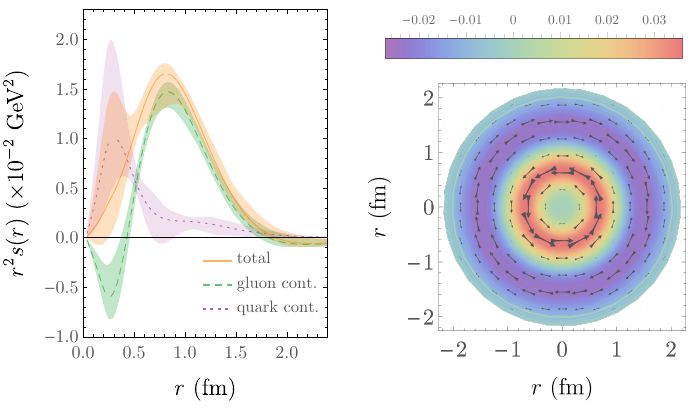}
	\caption{Left) Quark (purple) and gluon (green) shear distributions in the proton determined from modified $z$-expansion fits to the LQCD GFFs in the $\overline{\rm MS}$ scheme at $\mu=2$ GeV, as well as the total shear (orange) defined as their sum. Right) Tangential forces in the proton. The colour-coding and arrows represent the tangential shear vector field defined in Ref.~\cite{Polyakov:2018zvc}.
	\label{fig:shear}}
\end{figure}

 In Fig.~\ref{fig:shear}, the quark and gluon shear forces in the proton, determined from modified $z$-expansion fits to the $D$-term GFFs (Eq.~\eqref{eq:modz}) are shown, along with a rendering of the tangential forces in the proton \cite{Polyakov:2018zvc}. 

The shear and pressure distributions can be combined to define a mechanical radius of the proton~\cite{Polyakov:2018zvc},  $\langle r^2_{\rm Mech.} \rangle = \int r^2 Z(r) d^3r / \int Z(r) d^3r$ where $Z(r)=\frac{2}{3}s(r) + p(r)$. Using the pressure and shear distributions determined from the LQCD results, this is found to be $\langle r^2_{\rm Mech.} \rangle = 0.51(2)$ fm$^2$  using the modified $z$-expansion to parametrise the $D$-term GFFs and 0.57(1) fm$^2$ using the tripole ansatz. This is smaller than the experimentally determined charge radius of the proton, but similar to the charge radius calculated from LQCD  at heavier quark masses comparable to those used here \cite{Hasan:2017wwt}.  \\[-6pt]

{\bf Comparison to BEG $D$-term and future experimental goals:}
In Fig.~\ref{fig:comparison}, the BEG quark $D$-term form factor extracted from DVCS is compared with the LQCD determinations of the quark and gluon form factors. The BEG result has been shifted to the renormalisation scale $\mu=2$ GeV in the $\overline{\text{MS}}$ scheme using the three-loop running~\cite{Gracey:2003mr}\footnote{The result illustrated in Fig. 4 of BEG has  been rescaled by  18/25 to relate the DVCS extraction to the flavour-singlet combination, under the assumptions of BEG, and to match  Eq.~\eqref{Eq:EMT-FFs-spin-12} (BEG use an alternate notation  $d_{1,q}(t,\mu) = \frac{4}{5}D_q(t,\mu)$ and have $\mu^2=1.4$~GeV$^2$). The systematic uncertainties presented in the experimental extraction of $D_q(t,\mu)$ \cite{Burkert:2018bqq}  have been included in quadrature.}.
The connected isoscalar quark GFF determined from LQCD is  approximately $1.7\times$ smaller in magnitude than the BEG GFF, albeit with significant uncertainties.  
The LQCD determination of the gluon $D$-term form factor is noticeably larger in magnitude than the BEG result. It also favours a more general functional form in $t$ than the tripole assumed in BEG, although it is not inconsistent with a tripole ansatz within uncertainties. 

\begin{figure}[!t]
	\includegraphics[width=1.\columnwidth]{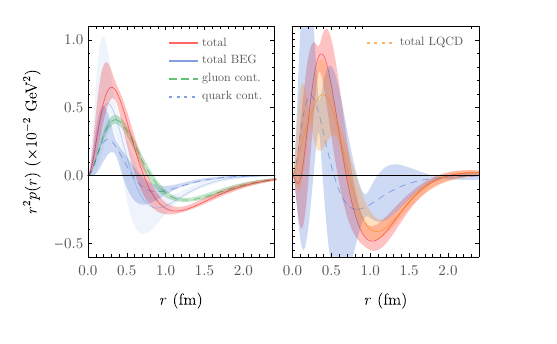}
	\caption{Left) Pressure distribution of the proton determined from tripole parametrisations of the BEG quark GFF and the LQCD gluon GFF. The red band corresponds to the total pressure distribution, while the dark blue dotted and green dashed bands denote to the quark and gluon contributions to the total. The pressure under the BEG assumption that that $D_g(t,\mu)=D_q(t,\mu)$ is shown as the blue solid band. Right) The same totals computed based on modified $z$-expansion fits to the GFFs. Also shown is the result obtained using only LQCD data, parametrised using the modified $z$-expansion (orange dashed band).
		\label{fig:BEGpressure}}
\end{figure}

The BEG analysis assumes that $D_{g}(t,\mu)=D_{q}(t,\mu)$  as there is no information on the gluon $D$-term from experiment. This is in mild tension with the LQCD results, and, moreover, given the scale evolution, Eq.~\eqref{eq:evolve}, can only possibly hold at one scale.  Since DVCS accesses the charge-squared weighted combination of quark flavours, BEG also assumes that the isovector quark contributions to the $D_q(t,\mu)$ form factor vanish, i.e., $D_u(t,\mu)=D_d(t,\mu)$. The LQCD finding that $D_{u-d}(t,\mu)\sim0$ provides compelling motivation for this assumption (large $N_c$ arguments \cite{Goeke:2001tz} also support it). 
The left panel of Fig.~\ref{fig:BEGpressure} shows the pressure distribution of the proton computed from the BEG quark $D$-term GFF and the LQCD gluon GFF, both parametrised using a tripole form and assuming that the quark-mass dependence of the latter is negligible in comparison with the statistical uncertainties. This pressure distribution is consistent within uncertainties with the determination using only LQCD data.
The pressure obtained under the assumptions of BEG (i.e., $D_g(t,\mu)=D_{u+d}(t,\mu)$) is also displayed. In comparison with the BEG assumption, the inclusion of the LQCD gluon contribution shifts the peaks of the pressure distribution outwards and extends the region over which the pressure is non-zero. 

As discussed above, the tripole form assumed for $D_q(t,\mu)$ in BEG introduces significant model-dependence into the pressure extraction (as detailed in the Supplementary Material, more general fit forms such as the modified $z$-parameter expansion with 3 parameters are not disfavoured by consideration of the Bayes Information Criterion).
 With the limited kinematic range of the CLAS data this is particularly problematic; the LQCD calculations show that the quark and gluon $D$-term GFFs have significant support up to $|t| \sim 2\ {\rm GeV}^2$ (assuming weak quark-mass dependence), which is far beyond the range of the experimental data. Fig.~\ref{fig:comparison} shows the result of a modified $z$-expansion fit to the BEG $D$-term form factor; outside the data range, the parametrisation is very poorly constrained. As shown in the right panel of Fig.~\ref{fig:BEGpressure}, this more general fit leads to a pressure distribution  that is consistent with zero everywhere within two standard deviations, demonstrating that experimental data over a larger kinematic range is needed before a model-independent extraction of the pressure is possible. 

In order to investigate the range of $t$ required for a model-independent pressure extraction from  experiment, fake data for the quark $D$-term GFF are generated in intervals of $\Delta t=0.1$ GeV$^2$ extending the experimental data along the tripole fit, assuming uncertainties of the same size as the average uncertainty in the BEG GFF  determination. The consistency of the LQCD data with a tripole form gives confidence that such an extension is justified. These fake data are then used to constrain a modified $z$-expansion fit and calculate the corresponding pressure distribution. For a determination of the pressure distribution that is distinct from zero at 2 standard deviations at the maximum of the first peak, the range of the experimental data must be extended in this manner to at least  $|t|\sim 1.0$ GeV$^2$. Future experiments, such as those using the CLAS12 detector at JLab and a future EIC, should seek to extend the kinematic reach to address this deficiency, even at the expense of precision in individual $t$ bins. With the EIC's potential \cite{Accardi:2012qut,Aschenauer:2013hhw} to determine the gluon GPDs that are  necessary in defining the pressure, similar kinematic coverage should be the goal of EIC experiments. Finally, the flavour separation necessary for a complete determination of the pressure distribution can be enabled by studies of deeply-virtual meson production and DVCS on deuterons \cite{Accardi:2012qut,Aschenauer:2013hhw}.\\[-6pt]

{\bf Summary:}
The shear and pressure distributions of the proton are determined from LQCD calculations for the first time. The results indicate that gluons play an important role in the internal dynamics of the proton, distinct from that of quarks. In particular, the gluon contributions to the $D$-term form factor, from which the pressure and shear distributions are defined, dominate the quark terms at the scale $\mu=2$~GeV in the $\overline{\text{MS}}$ scheme.
These calculations are undertaken at heavier-than-physical quark masses corresponding to a pion mass roughly three times the physical value, and at a single lattice spacing and volume. LQCD calculations at the physical quark mass, in multiple volumes and with multiple lattice spacings, and which include the effects of quark and gluon operator mixing and disconnected quark contributions, offer the prospect of a controlled, and model-independent, theoretical determination of the shear and pressure distributions of the proton. With improved LQCD algorithms and growing computational resources, this goal is eminently feasible and will set important benchmarks for measurements using the CLAS12 detector at JLab and at a future EIC.

This study provides support for some of the  assumptions made in the recent first extraction of the pressure distribution of the proton from DVCS experiments at JLab. However, given the strong model-dependence involved in the relation of the $D$-term form factor to the shear and pressure distributions, it is found that a clean experimental determination of these quantities will require  flavour-separated measurements of the quark $D$-term form factors over a kinematic range extending over at least $0\le |t| \alt 1$ GeV$^2$, as well as constraints on the gluon $D$-term form factors for similar kinematics.

\section*{Acknowledgements}
The authors thank Volker Burkert,  Ross Young and James Zanotti for helpful discussions and comments.
This work used the facilities of the Extreme Science and Engineering Discovery Environment (XSEDE), which is supported by National Science Foundation grant number ACI-1548562, under allocation TG-PHY170018, as well as facilities of the USQCD Collaboration, which are funded by the Office of Science of the U.S. Department of Energy. Resources of the National Energy Research Scientific Computing Center (NERSC), a U.S. Department of Energy Office of Science User Facility operated under Contract No. DE-AC02-05CH11231, were also used, as were computing facilities at the College of William and Mary which were provided by contributions from the National Science Foundation, the Commonwealth of Virginia Equipment Trust Fund and the Office of Naval Research.
PES is supported by the National Science Foundation under CAREER Award 1841699 and in part by Perimeter Institute for Theoretical Physics. Research at Perimeter Institute is supported by the Government of Canada through the Department of Innovation, Science and Economic Development and by the Province of Ontario through the Ministry of Research and Innovation. WD is supported by the U.S. Department of Energy under Early Career Research Award DE-SC0010495 and grant DE-SC0011090. WD is also supported  within the framework of the TMD Topical Collaboration of  the U.S. Department of Energy, Office of Science, Office of Nuclear Physics, and  by the SciDAC4 award DE-SC0018121.

\bibliography{bob}

\appendix

\end{document}